\documentclass[12pt]{iopart}
\usepackage{graphicx}

\begin{document}

\title[Confined nano-droplet]{Phase behavior of a confined nano-droplet in the grand-canonical ensemble: the reverse liquid-vapor transition.}
\author{James F. Lutsko}
\address{Center for Nonlinear Phenomena and Complex Systems CP 231, Universit\'{e}
Libre de Bruxelles, Blvd. du Triomphe, 1050 Brussels, Belgium}
\ead{jlutsko@ulb.ac.be}
\author{Julien Laidet}
\address{Ecole Normale Supérieure, D\'{e}partement de Physique, 24 rue Lhomond, 75231 Paris Cedex 05}
\author{Patrick Grosfils}
\address{Microgravity Research Center, Chimie Physique E.P. CP 165/62, Universit\'{e}
Libre de Bruxelles, Av.F.D.Roosevelt 50, 1050 Brussels, Belgium.}

\begin{abstract}
The equilibrium density distribution and thermodynamic properties of a Lennard-Jones fluid confined to nano-sized spherical cavities at constant chemical potential was determined using Monte Carlo simulations. The results describe both a single cavity with semipermeable walls as well as  a collection of closed cavities formed at constant chemical potential. The results are compared to calculations using classical Density Functional Theory (DFT). It is found that the DFT calculations give a quantitatively accurate description of the pressure and structure of the fluid. Both theory and simulation show the presence of a ``reverse'' liquid-vapor transition whereby the equilibrium state is a liquid at large volumes but becomes a vapor at small volumes.
\end{abstract}

\date{\today }


\submitto{\JPCM}

\maketitle

\section{Introduction}

The current intense interest in nano-scale systems provides strong motivation for developing simple means to predict the properties of small systems. One possible approach to this problem is the use of quantitatively accurate classical Density Functional Theory. Classical DFT has long been used to study the properties of bulk liquid-vapor interfaces, solids and liquids in confined geometries such as slit-pores and near walls\cite{FundInhomLiq,Wu,bonn:739,lutsko_jcp_2008_2}. However, there have been few quantitative tests of the theories for truly small systems consisting of dozens to hundreds of atoms. In this paper, we present one such test in which DFT calculations are compared to simulation for the case of a liquid confined to a small spherical cavity. 

Density Functional Theory is most easily formulated in the grand-canonical ensemble\cite{EvansDFT,HansenMcdonald, LutskoRev, Gonzalez}. It can be applied to other ensembles, but this requires further expansions and approximations\cite{talanquer:5190,Gonzalez,Kosov}. In the thermodynamic limit, the difference between the ensembles is of little practical importance. However, for finite systems - especially small finite systems - the difference between the ensembles becomes qualitative\cite{Hill, Gonzalez}. For these reasons, we have chosen to work in the grand-canonical ensemble where comparisons can be made with the fewest assumptions. Physically, a finite-volume system in the grand canonical ensemble is not without interest as it describes  a single cavity with a hard, but semi-permeable wall or the average properties of a collection of cavities of the same size but with different numbers of  particles\cite{Gonzalez}. 

We note that other approaches to the description of confined fluids exist. In particular, integral equation methods from liquid state theory have been used to study the structure and thermodynamics of charged fluids in a charged spherical pore\cite{PhysRevLett.79.3656} as well as that of hard-sphere fluids in slit and cylindrical pores\cite{Alejandre}. The latter work compares the results of the calculations to simulations in the grand ensemble and is therefore complementary to the present study. 

In the following, we compare the results of Monte Carlo simulations and DFT calculations performed in the grand-canonical ensemble for a system consisting of point atoms interacting via a Lennard-Jones potential and confined to a spherical cavity by hard walls. In both the simulations and the DFT calculations, the walls are instantiated by an applied field which is zero for particles inside the cavity and large (tending to infinity) for particles outside the cavity. Thus, the variables characterizing the state of the system are temperature, chemical potential and the size of the cavity. We find that for a value of the chemical potential corresponding to a stable liquid phase in the bulk system, and a metastable vapor phase, the system undergoes a ``phase transition'' as the volume is reduced whereby the vapor phase becomes the preferred state for small volumes. Of course, in finite system, we do not observe a true phase transition in the thermodynamic sense, but rather a hysteresis in the simulations. The calculations, since they yield a free energy, do allow us to specify the location of the transition in the sense of the volume at which the free energies are equal. 

In the next Section, we briefly describe our simulation technique and our calculations. The calculations are performed using the Modified-Core Van der Waals (MC-VDW) model DFT\cite{lutsko_jcp_2008_2}. This model is based solely on properties of the bulk fluid and the interaction potential and gives a quantitatively accurate description of the fluid under a wide variety of pair potentials and external fields. A comparison between theory and simulation is presented in the third Section where a transition between the vapor phase at small volumes and the liquid phase at large volumes is described. Our conclusions are summarized in the final Section.

\section{Simulation and Calculations}

\subsection{Simulations}

We have carried out simulations in the grand canonical ensemble of particles of mass $m$ and positions $\mathbf{r}_i$ and momenta $\mathbf{p}_i$. The N-particle Hamiltonian is
\begin{equation}
H=\sum_{i=1}^{N}\frac{p_i^2}{2m}+\sum_{i<j}v(r_{ij})+\sum_{i=1}^{N}\phi(r_{i})
\end{equation}
where the pair potential is the Lennard-Jones interaction,
\begin{equation}
v(r)=4\epsilon\left(\left(\frac{\sigma}{r}\right)^{12}-\left(\frac{\sigma}{r}\right)^{6}\right)
\end{equation}
and where the external field, $\phi(r)$ is taken to be zero for $r<R$
and infinite for $r>R$. Since the available volume is finite, no truncation of the potential is necessary. Our simulations follow the procedure described
in ref.\cite{FrenkelSmit}. Each simulation consists of a large number of ``cycles'' consisting  of $0.9N$ attempts to move
a particle together with $0.05N$ attempts to add a particle and
$0.05N$ attempts to remove a particle for a total of $N$ attempted
changes. Particle moves consist of choosing a random vector of maximum
length $\Delta$ which is added to the coordinates of a randomly chosen
particle. The move is then evaluated using the usual Metropolis
algorithm\cite{FrenkelSmit}. The effect of the external field is that all moves resulting in
particles being outside the spherical cavity of radius $R$ are
rejected. Insertions consist of adding a particle at a random position
within a cube with sides of length $2R$ and accepting or rejecting
according to the Metropolis algorithm based on the total energy
$E=H-\mu N$ where $\mu$ is the imposed chemical potential. Deletions
are attempted in the obvious way. The parameter $\Delta$ is chosen to
give an acceptance rate on the order of $50\%$. 

Note that in our simulations, the particles are treated as points relative to the boundary. Our results apply equally well to particles that behave as hard spheres when interacting with the wall. If the hard sphere diameter is $d$ then the properties of the system with cavity radius $R_d$ will correspond to one with $d=0$ and cavity radius $R=R_d-d/2$.  

The simulations begin with a random distribution of particles that is allowed to equilibrate for several million cycles. Then, statistics including the average number of particles, total energy, virial pressure and density profile are accumulated over a run of 5 million cycles. The density profiles were calculated by tracking the number of particles in equal volume shells both relative to the center of the cavity and relative to the center of mass. We found little difference in the two profiles and report the former here.

\subsection{Density Functional Theory}

In DFT, the properties of the system are expressed in terms of the local density $\rho \left( \mathbf{r} \right)$. For example, the average number of particles is $<N> = \int \rho \left( \mathbf{r} \right) d\mathbf{r}$. The equilibrium density is determined by minimizing the functional $\Omega[\rho] \equiv F[\rho] - mu <N>$ where $F[\rho]$ plays the role of the Helmholtz free energy and is not, in general, known exactly. The equilibrium grand potential is then equal to $\Omega[\rho]$ evaluated at the equilibrium density. Our DFT calculations were performed using the MC-VDW model\cite{lutsko_jcp_2008_2}. This model is an extension of the simplest hard-core plus mean-field tail model which gives quantitatively accurate predictions for surface tension\cite{lutsko_jcp_2008_2}, fluid structure in slit pores\cite{lutsko_jcp_2008_2}, nucleation barriers\cite{lutsko_jcp_2008_3}, etc. Since the aim here is to compare directly to simulation, quantitative accuracy of the DFT calculations is a necessity. The model is written as a sum of four contributions,
\begin{equation}
F[\rho] = F_{id}[\rho] + F_{hs}[\rho] + F_{core}[\rho] + F_{tail}[\rho].
\end{equation}
The first contribution is the ideal gas term which is given by
\begin{equation}
F_{id}[\rho] = \int \left( \rho(\mathbf{r}) \log\left( \rho(\mathbf{r}) \right) -\rho(\mathbf{r}) \right) d\mathbf{r}.
\end{equation}
Next is a hard-sphere contribution, $F_{hs}[\rho]$,
for which the ``White Bear'' Fundamental Measure Theory (FMT) model was
used\cite{WhiteBear, tarazona_2002_1} along with the Barker-Henderson
hard-sphere diameter\cite{BarkerHend,HansenMcdonald}. The third contribution,
the ``core correction'' $F_{core}[\rho]$,
is similar to a FMT model but is constructed so that the total free
energy functional reproduces a given equation of state in the bulk
phase as well as certain other conditions concerning the direct
correlation function in the bulk fluid\cite{lutsko_jcp_2008_2}. The
final term is a mean-field treatment of the long-range attraction,
\begin{equation}
F_{tail}[\rho] = \int \Theta(r_{12}-d) \rho(\mathbf{r}_{1})
\rho(\mathbf{r}_{2})v(r_{12}) d\mathbf{r}_{1} d\mathbf{r}_{2},
\end{equation}
where $\Theta(x)$ is the step function, $d$ is the Barker-Henderson
hard-sphere diameter and $v(r)$ is the pair potential. The DFT model requires as input the \emph{bulk} equation of state. Since the object of the
calculations was to model the LJ system as accurately as possible, the
empirical equation of state of Johnson, Zollweg and Gubbins
\cite{JZG} was used. 

The DFT calculations were performed assuming a spherically symmetric density profile which was discretized as a function of distance from the center, $r$ with $160$ points per hard-sphere diameter. This rather fine grid was necessary so as to minimize the discretization effects at the discontinuity at the boundary of the cavity.

\subsection{Theory: Exact results}
For  small volumes, it is very unlikely that there will be more than one or two particles present due to the divergent repulsion at small distances. In this case, the grand partition function can be approximated by
\begin{eqnarray}
\Xi &=&\sum_{N=0}^{\infty }\exp \left( \beta \mu N\right) Z_{N} \label{exact}\\
&=&1+\exp \left( \beta \mu \right) Z_{1}+\exp \left( 2\beta \mu \right)
Z_{2}+...  \nonumber \\
&=&1+zZ_{1}+z^{2}Z_{2}+...  \nonumber
\end{eqnarray}%
where $Z_N$ is the canonical partition function for a system of $N$ particles. For the cases $N=1,2$ straightforward calculation taking into account the finite volume of radius $R$ gives
\begin{eqnarray}
Z_{1} &=&\Lambda ^{-3}V \\
Z_{2} &=&\frac{1}{2}\Lambda ^{-6}V\frac{\pi }{4}\int_{0}^{2R}\left( 16-12%
\frac{r}{R}+\left( \frac{r}{R}\right) ^{3}\right) r^{2}\exp \left( -\beta
v\left( r\right) \right) dr  \nonumber
\end{eqnarray}%
with the thermal wavelength
\begin{equation}
\Lambda = \sqrt{\frac{h^{2}}{2\pi mk_{B}T}}
\end{equation}%
From these expressions, the grand potential, $\Omega = -k_{B}Tln\Xi$ can be calculated and thermodynamic properties such as the average number of particles, the pressure, etc. determined by differentiation. This result gives a further check on the DFT calculations as well as a consistency check for the simulations.

\section{Results}
\begin{figure*}[htpb]
\begin{center}
\resizebox{12cm}{!}{
{\includegraphics[height = 8cm, angle=0]{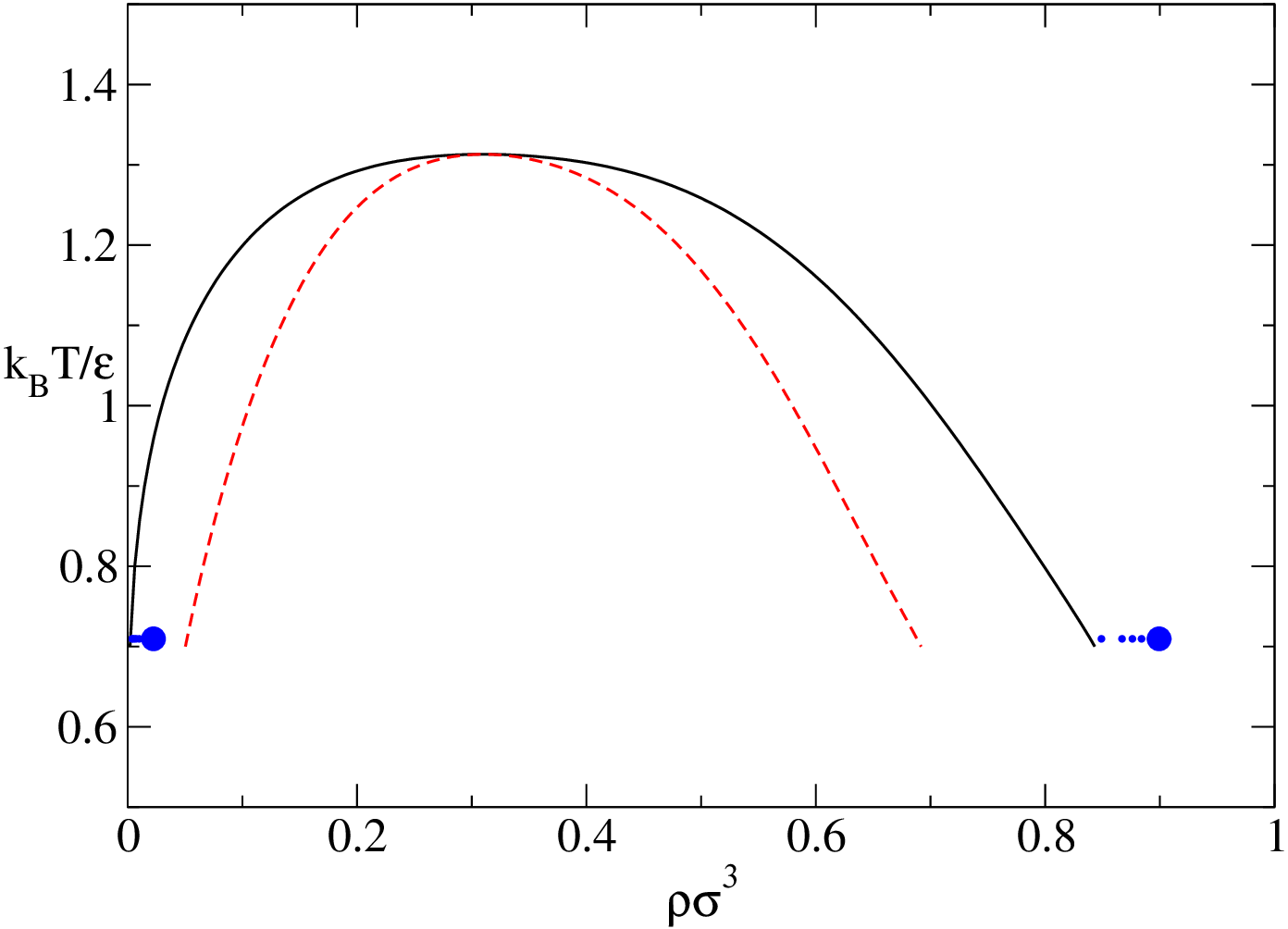}}}
\end{center}
\caption{(Color on line) The phase diagram of the LJ fluid as calculated using the JZG equation of state. The full line is the binodal and the dashed line the spinodal. The large spots correspond to chemical potential $\mu=-3$ and the smaller spots are the states sampled when the chemical potential is varied.}
\label{fig1}
\end{figure*}

In the following, we take $\epsilon$ and $\sigma$ to be the units of energy and length, respectively, so all quantities can be considered to be dimensionless. Figure \ref{fig1} shows the bulk phase diagram of the Lennard-Jones fluid with the thermodynamic states investigated here indicated. We work at a temperature of $k_BT=0.71\epsilon$ which is approximately the triple point of the LJ potential. In the first set of investigations, the volume is varied with the chemical potential  fixed at a value of $\mu = -3 \epsilon$ corresponding to a stable liquid with density $\rho\sigma^3=0.899$ and a vapor in the metastable region. In the second set of investigations, the chemical potential is varied so as to move the liquid phase towards the binodal (i.e. decreasing chemical potential) and the volume is held fixed. 

\subsection{Variation of volume}
\begin{figure*}[htpb]
\begin{center}
\resizebox{12cm}{!}{
{\includegraphics[height = 8cm, angle=0]{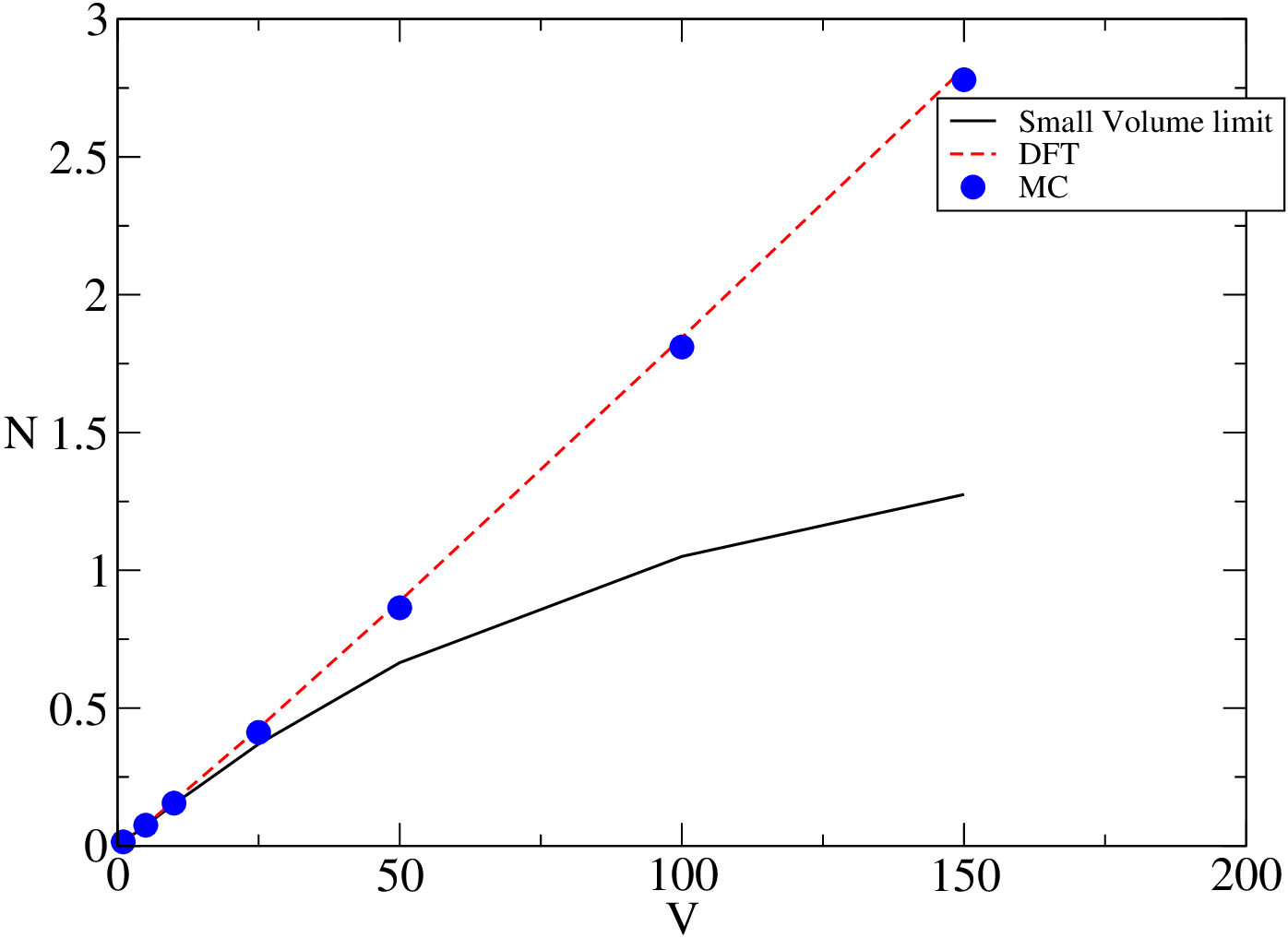}}}
\end{center}
\caption{(Color on line) The average number of particles as a function of volume at fixed chemical potential $\mu = -3.0$ and temperature $k_BT=0.71$ as determined from Eq.( \ref{exact}), simulation and DFT.}
\label{fig2}
\end{figure*}

\begin{figure*}[htpb]
\begin{center}
\resizebox{12cm}{!}{
{\includegraphics[height = 8cm, angle=0]{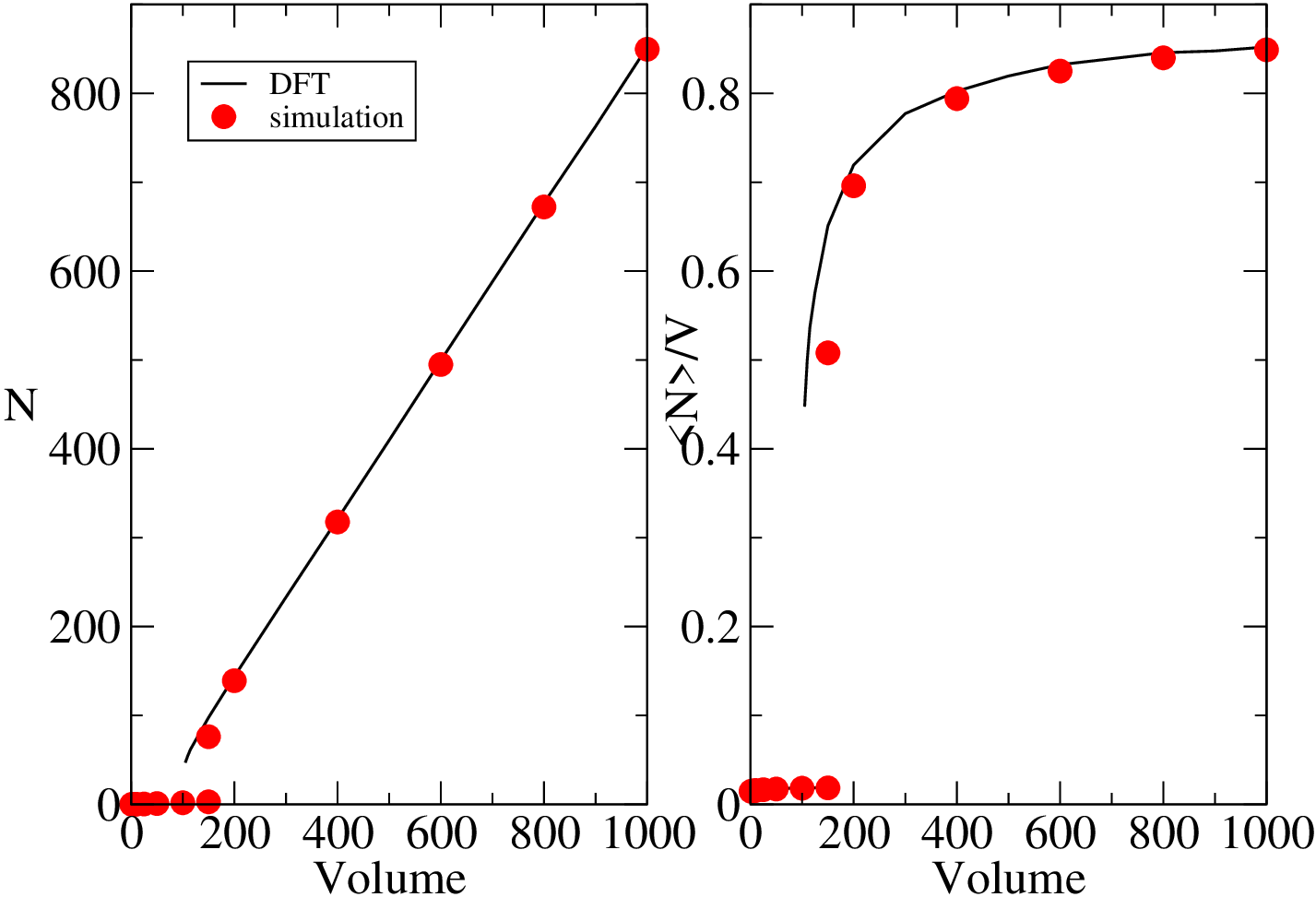}}}
\end{center}
\caption{(Color on line) The left panel shows the average number of particles as a function of volume at fixed chemical potential $\mu = -3.0$ and temperature $k_BT=0.71$ as determined from simulation and DFT. The panel on the right shows the average density as a function of volume.}
\label{fig3}
\end{figure*}

We now consider the variation of the volume at a constant chemical potential, $\mu=-3.0$. Figure \ref{fig2} shows the number of particles as a function of volume for small volumes as determined by simulation, DFT and via the usual thermodynamic relation
\begin{equation}
<N> = \frac{\partial \Omega}{\partial \mu}
\end{equation}
using the small volume approximation given in Eq.(\ref{exact}). The DFT calculations are in good agreement with the simulations and both approach the analytic small volume limit for $V \leq 5$. 
Figure \ref{fig3} shows the average number of atoms and the density for a wide range of volumes as determined from  both DFT calculations and simulation. As seen in the figures, there are two phases possible, depending on the volume: at low volumes, the system is always a low-density gas while at high volumes it is always a high-density liquid. This is therefore the inverse of the expected behavior in a canonical ensemble where we expect a condensed phase to occur at low volumes and a gas at high volumes. In the present case, this is not a true thermodynamic phase transition because of the finite size of the systems, so at intermediate volumes both phases are stable over the time-scale of the simulation.The same behavior is observed in the calculations where it is possible to stabilize both phases for $100 < V < 200$ (the ``two phase region'') while otherwise, only the liquid (vapor) is stable at higher (lower) volumes. 

\begin{figure*}[htpb]
\begin{center}
\resizebox{12cm}{!}{
{\includegraphics[height = 8cm, angle=0]{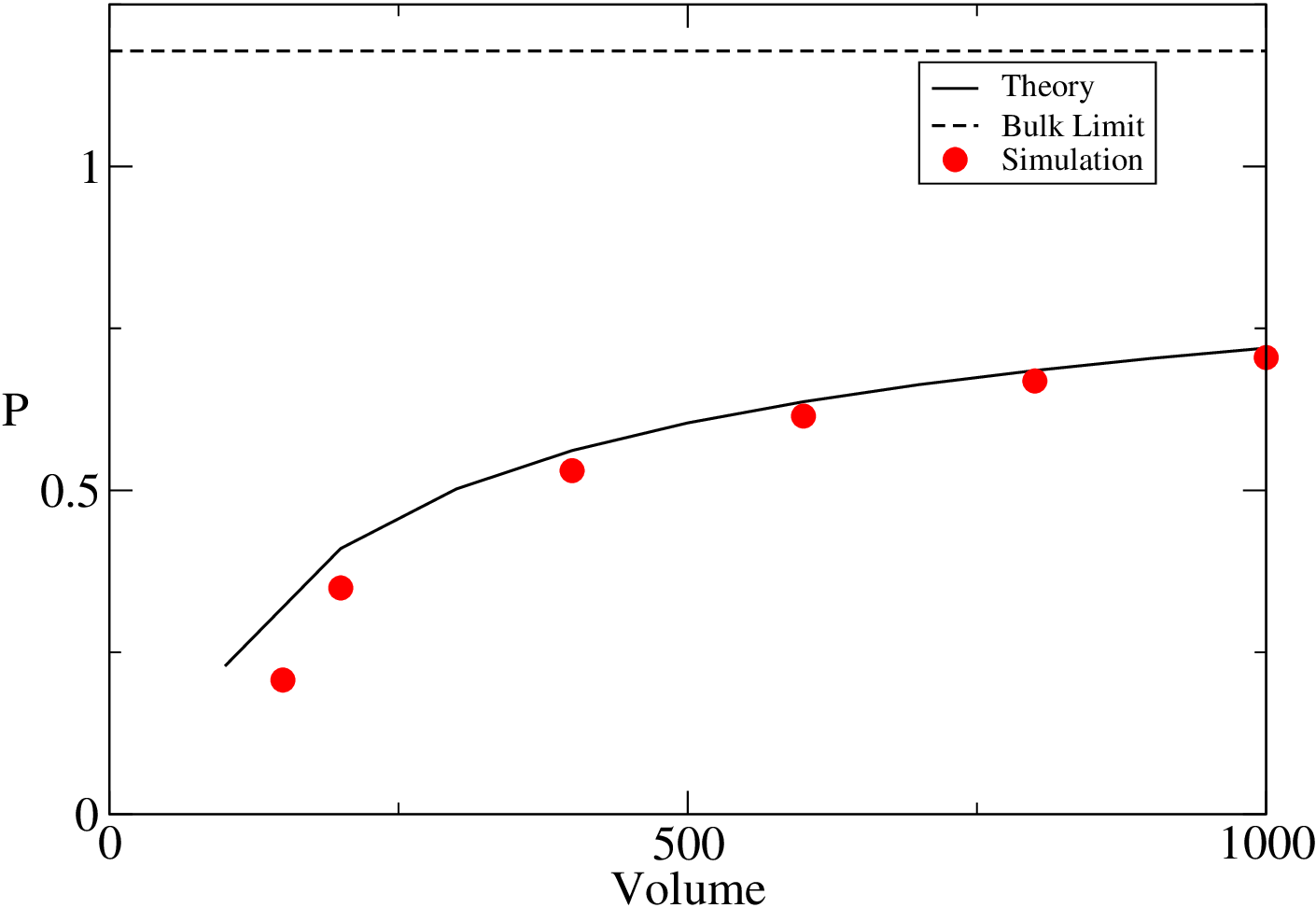}}}
\end{center}
\caption{(Color on line) The pressure as a function of volume at fixed chemical potential $\mu = -3.0$ and temperature $k_BT=0.71$ as determined from simulation and DFT.}
\label{fig4}
\end{figure*}

\begin{figure*}[htpb]
\begin{center}
\resizebox{12cm}{!}{
{\includegraphics[height = 8cm, angle=0]{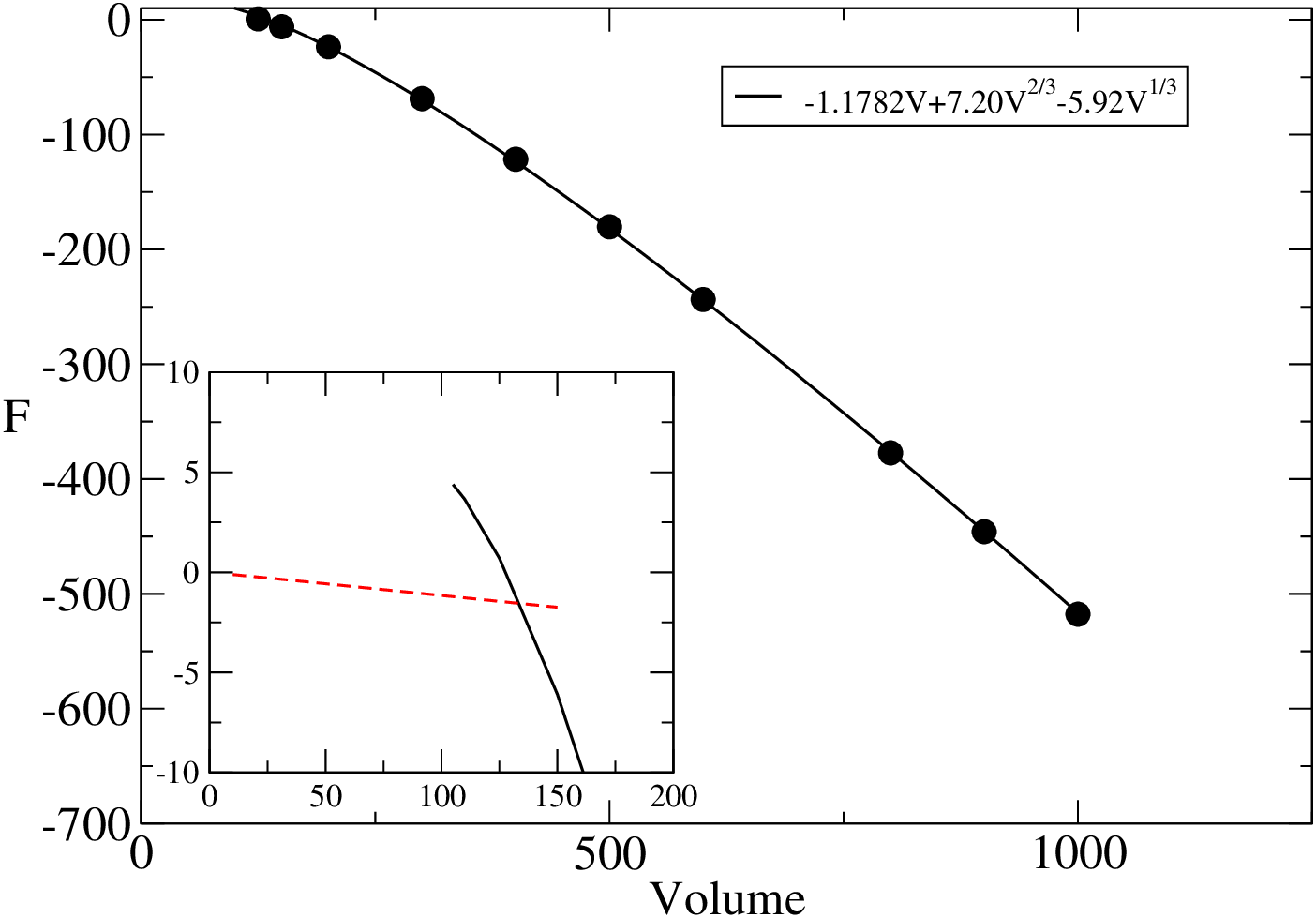}}}
\end{center}
\caption{(Color on line) Free energies of the liquid and gas phases as functions of the volume at fixed chemical potential $\mu = -3.0$ and temperature $k_BT=0.71$ as determined from DFT. In the main figure, the calculated values are shown as symbols and the best fit to a function of the form $F=aV+bV^{2/3}+cV^{1/3}$ is shown as the full line (where the first coefficient, $a$, is fixed by the bulk limit). The inset shows that the free energies of the liquid (solid line)  and vapor (dashed line) phases are equal at about $V=132$.}
\label{fig5}
\end{figure*}

Figure \ref{fig4} shows the pressure ($P=-\partial \Omega / \partial V$) where the agreement between DFT and simulation is again quite good. At the largest volumes shown, the pressure is still far below the bulk limit. As the volume decreases towards the two phase region, there is a sharp drop in pressure and it is here that the largest differences between DFT and simulation occur. The free energies of both phases, as determined from the calculations, is shown in Fig. \ref{fig5} where the crossover occurs at $V \sim 132$. Figure \ref{fig5} also shows that at large volumes, the free energy has the expected form of a bulk contribution, linear in the volume, and a surface term that varies as $V^{2/3}$. It is the surface term that gives rise to a very slow $V^{-1/3}$ convergence of the pressure to the bulk limit, as is seen in Fig. \ref{fig4}. In fact, fitting the simulation data for the pressure for $V > 200$ to the function $P = a + b*V^{-1/3}$ gives and estimate $a=1.196$ for the bulk pressure which is very close to the value of $p_{0}=1.178$ given by the JZG equation of state. 

Based on this behaviour, the observed transition can be understood with a simple capillary model. For a sufficiently large system, the free energy will consist of two contributions: the free energy of the gas far from the wall, which will be in the bulk state, and a contribution from the interaction between the fluid and the wall. The latter has the effect of a surface tension so that, in the simplest, capillary approximation, the grand potential of the fluid will be 
\begin{equation}
\beta \Omega = \frac{4\pi}{3}R^{3}\left( f(\rho)-\mu \rho\right)+4\pi R^{2} l \rho \tau
\end{equation}
where $\rho$ is the average density, $f(\rho)$ is the bulk-phase Helmholtz free energy per unit volume, $\tau$ represents the excess free energy per particle due to the interaction with the wall and $l$ is the penetration depth of the effect of the wall. This corresponds to the empirical variation with radius observed above. Minimizing this with respect to the density gives
\begin{equation}
\frac{df(\rho)}{d\rho} = \mu - \frac{3l}{R}  \tau
\end{equation}
For large cavities, the second term on the right is negligable and this simply says that the density is that of a bulk fluid at chemical potential $\mu$ (which picks out the liquid phase for the chemical potential used here). The effect of the wall is to shift the chemical potential to lower values until for sufficiently small $R$, the effective chemical potential favors the vapor phase thus giving rise to the transition. (Note that a more realistic model would include the density-dependence of $\tau$ but we do not expect this to give rise to any qualitative differences.)

\begin{figure*}[htpb]
\begin{center}
\resizebox{12cm}{!}{
{\includegraphics[height = 8cm, angle=0]{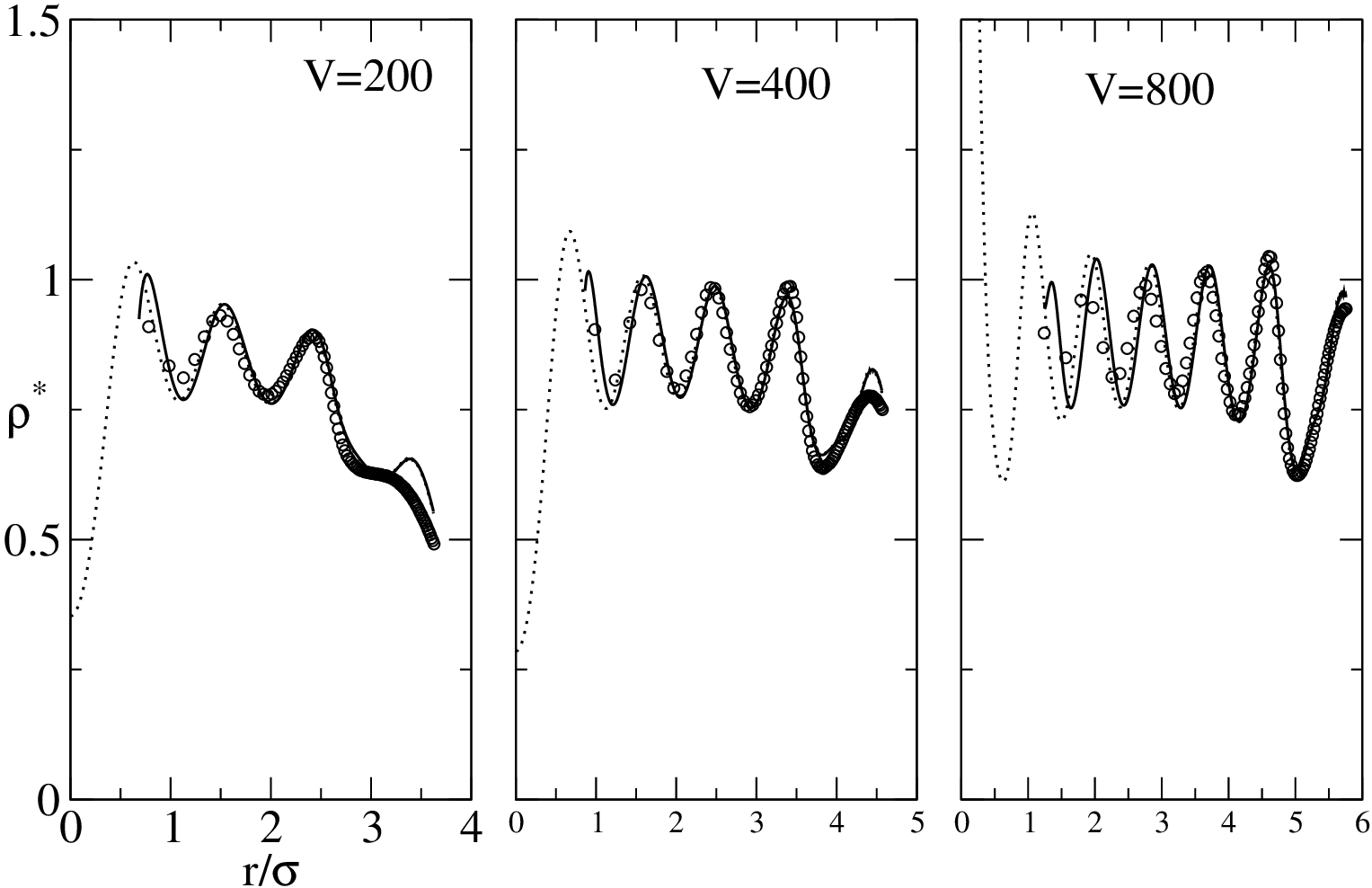}}}
\end{center}
\caption{The density profile for $V=200, 400$ and $800$  at fixed chemical potential $\mu = -3.0$ and temperature $k_BT=0.71$ as determined from simulation and DFT. The density profile from simulation (the circles) is calculated using 150 equal-volume shells. The DFT calculation is shown as the dotted line and the average of the DFT calculation over equal volume shells is shown as the thick line. Note that in each panel, the wall of the cavity corresponds to the right-most data point.}
\label{fig6}
\end{figure*}

As a further test of the ability of DFT to accurately described such small systems, we show in Fig. \ref{fig6} some examples of density profiles determined in the simulations compared to those calculated from DFT. For all of these systems, the fluid exhibits a shell structure which is accurately predicted by the DFT. In some cases, the density in the center of the cavity is very high (see third panel of \ref{fig6}) but this simply indicates a high \emph{probability density} of an atom occupying the center of the cavity and the physical quantity, which is the average number of atoms in a volume of given radius about the origin, is always finite. The greatest error appear near the wall where the DFT tends to over-estimate the density. Since the particles interact with the wall as hard points - i.e. as ideal gas particles - the pressure exerted on the system by the wall must be the same as it would exert on an ideal gas at the same density (i.e. the density of the real fluid adjacent to the wall). The role of the wall is to confine the fluid which means, if the pressure is positive, to balance the pressure so we conclude that the fluid pressure must be equal to that of an ideal gas at the density of the fluid at the wall ($P = \rho(R) k_{B}T$). For planar interfaces, this is called the ``wall theorem''\cite{Swol,FundInhomLiq}. Taking into account that what is measured in the simulation is the density in a small shell near the wall, and not the actual density \emph{at} the wall, this relation is in fact confirmed in the simulations. For example for $V=800$, the density at the wall is found to be $\rho(R)=0.944$ and the prediction $P = 0.944*0.71 = 0.67$ is consistent with the virial pressure which is found to be $0.669$. The discrepancy near the wall can therefore be traced to the overestimate of the pressure by the DFT as is seen in Fig. \ref{fig4}.

\begin{figure*}[htpb]
\begin{center}
\resizebox{12cm}{!}{
{\includegraphics[height = 8cm, angle=0]{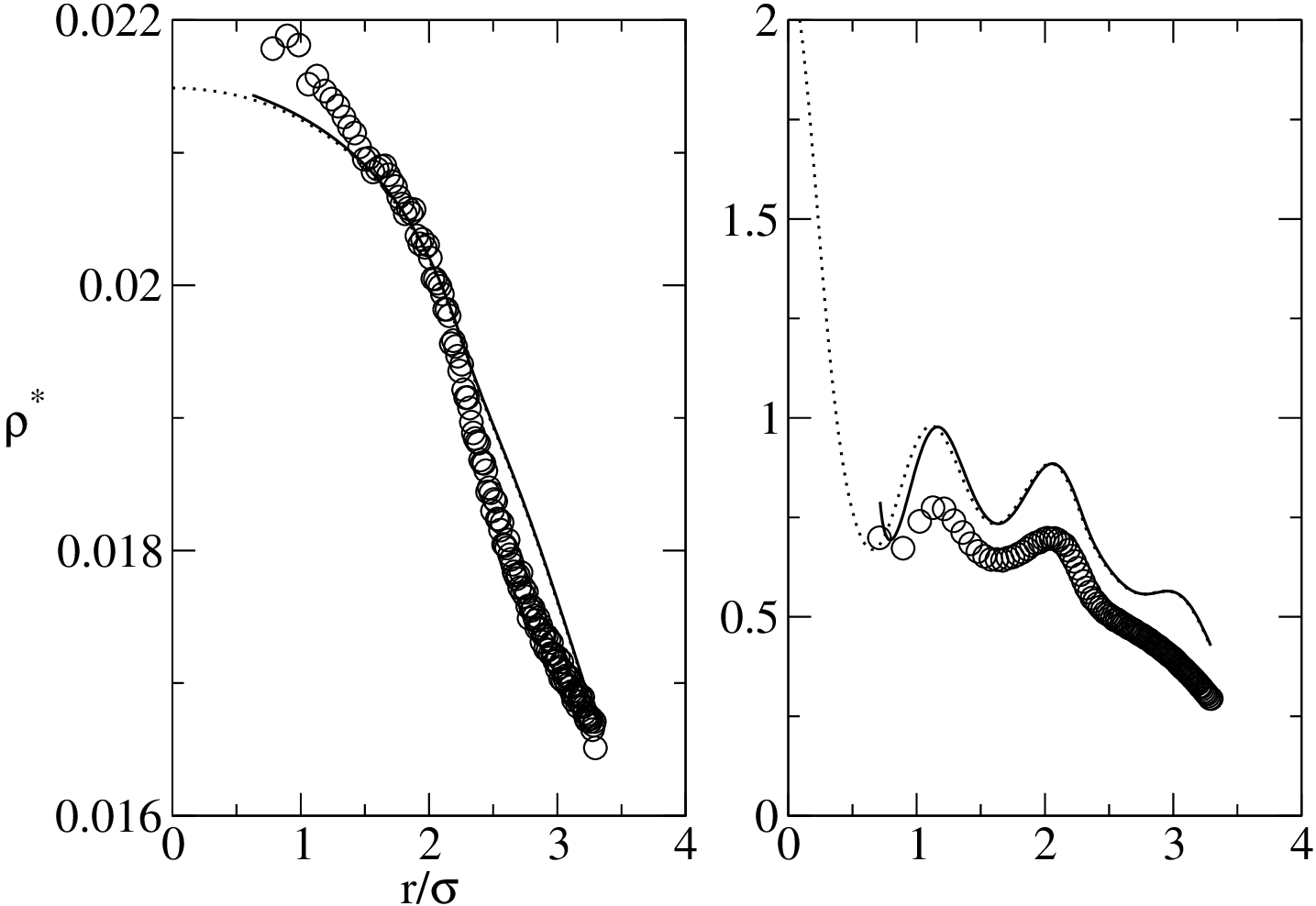}}}
\end{center}
\caption{The same as Fig. \ref{fig6} for $V=150$. Panel (a) shows the density distribution in the vapor phase and panel (b) shows the density distribution in the liquid phase.}
\label{fig7}
\end{figure*}

Figure \ref{fig7} shows the density in the metastable region (the case $V=150$) for both the vapor and liquid phases. In the vapor phase, the DFT calculations are in reasonable agreement with the simulations (away from the metastable region, agreement in the vapor phase is even better) but in the liquid phase the DFT is less accurate than elsewhere. This accords with the thermodynamic properties shown previously, which vary rapidly with volume and deviate most strongly from the DFT calculations in the metastable region and can be attributed to a small error in predicting the precise location of the ``phase transition''. 

\subsection{Variation of chemical potential}

We have also performed simulations and calculations at fixed volume ($V=800$) and temperature ($k_BT=0.71$) and with varying chemical potential. The results are briefly summarized here.

\begin{figure*}[htpb]
\begin{center}
\resizebox{12cm}{!}{
{\includegraphics[height = 8cm, angle=0]{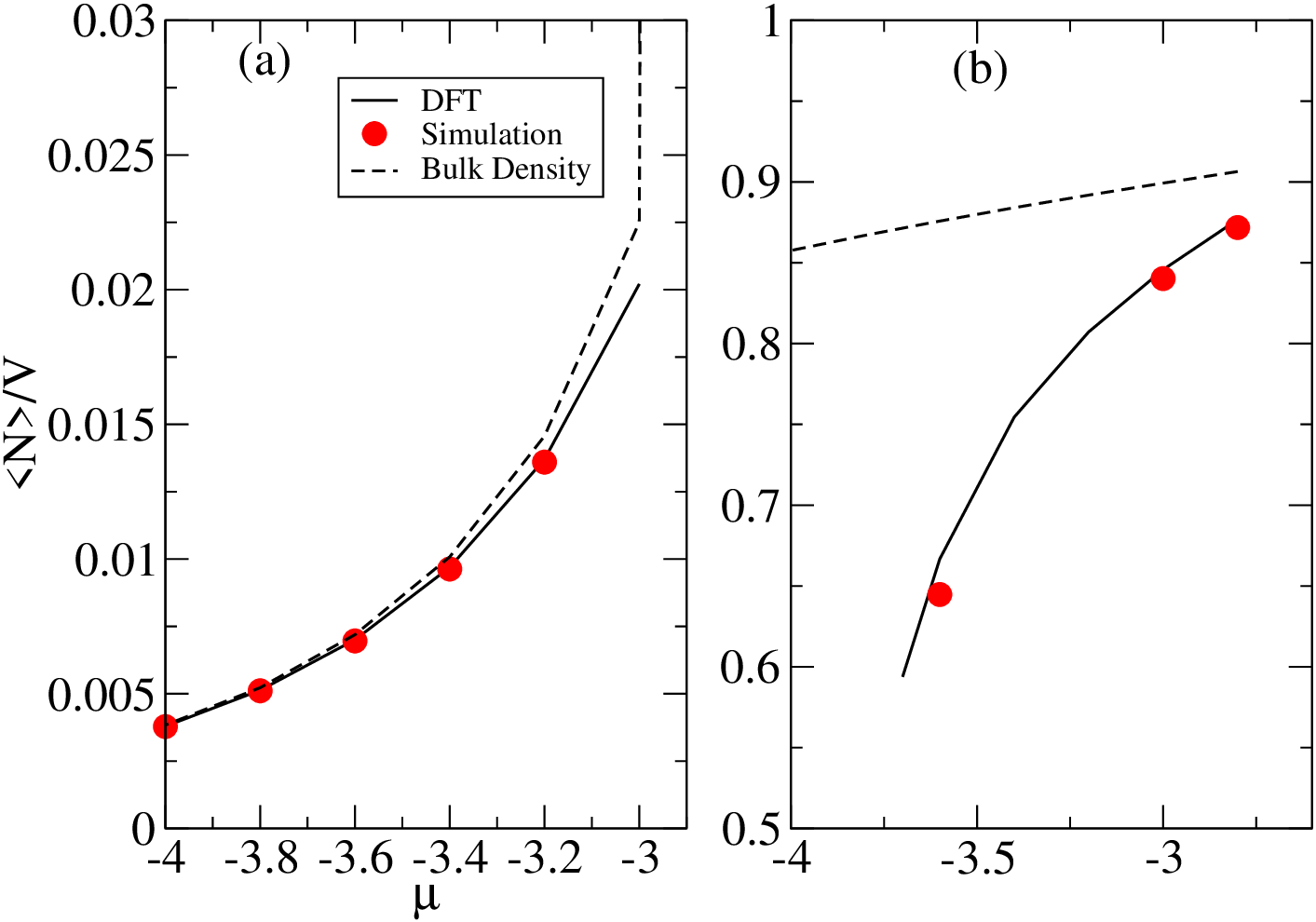}}}
\end{center}
\caption{(Color on line) The average density in the vapor (Panel a) and liquid (Panel b) phases as a function of chemical potential at $V=800$ and $k_BT=0.71$.}
\label{fig8}
\end{figure*}

Figure \ref{fig8} shows the liquid and vapor densities as function of the chemical potential. At very low chemical potential, the vapor is the stable phase and at higher chemical potentials, the liquid is the stable phase. A transition occurs at intermediate chemical potentials as signaled by the rapid drop in the average liquid density. DFT calculations of the free energies of the two phases indicate a transition at $\mu = -3.6$ which is consistent with the observed behavior in the simulations. A comparison of the density profiles is similar to that found at constant chemical potential: the DFT works well in both phases with the largest errors occurring for values of the chemical potential near the transition region. 

\section{Conclusions}
In summary, we have performed Monte Carlo simulations and DFT calculations of the thermodynamic properties and density profiles of a Lennard-Jones liquid confined to a spherical cavity with hard walls. At fixed chemical potential, we find a ``reverse'' liquid-vapor transition whereby the vapor is the stable phase at small volumes and the liquid is the stable phase at large volumes. Since the chemical potential corresponds to that of a stable liquid in the bulk limit, it is expected that the liquid is the stable phase at large, although finite, volumes. For any cavity, the particles near the wall have fewer neighbors than to particles in the bulk giving rise to a surface tension (or, more precisely, a surface excess free energy) as evidenced, e.g., by the fact that the free energy is well described by a function of the form $F=aV+BV^{2/3}$ at large volume. For small volumes, this surface tension dominates (i.e. a significant fraction of the system has lower coordination than in the bulk) so that the free energy is driven up until it exceeds that of the vapor (which is dominated by entropy and little affected by the boundaries). This competition between bulk and surface effects is completely analogous to the physics underlying classical nucleation theory (CNT). Thus, the instability of the liquid at small volumes is analogous to the instability of sub-critical clusters in CNT. Varying the chemical potential at fixed volume produces a standard liquid-vapor transition whereby the vapor is stable at very negative chemical potentials and the liquid at larger chemical potentials. 

Finally, one question motivating this study was whether DFT, which is based on properties of the bulk systems, is sufficiently versatile so as to be useful in predicting the properties of small, nano-scale systems. The answer is clearly affirmative for the particular model (MC-VDW) used here, with DFT giving a good description of the average (thermodynamic) properties as well as quantitatively reasonable predictions for the density distributions within the cavities. Since this model has been shown to work for a variety of semi-infinite systems\cite{lutsko_jcp_2008_2, lutsko_jcp_2008_3} as well as for different potentials\cite{Grosfils}, it is likely to be a useful tool in understanding the properties of more relevant nano-systems such as micro-plasmas and fluids in small pores and cavities in the canonical ensemble.

\ack
The work of JFL and JL was supported in part by the European Space Agency under contract number
ESA AO-2004-070. The work of PG was supported by the project ARCHIMEDES of the Communaut\'e
Fran\c caise de Belgique (ARC 2004-09). 

\bibliographystyle{unsrt}
\bibliography{liquid_vapor}

\end{document}